\documentclass[structabstract]{aa}

\usepackage{graphicx}
\usepackage{amssymb}

\usepackage{natbib}

\usepackage{txfonts}

\begin{document}

\title{Cosmography by gamma ray bursts}

\author{S. Capozziello\inst{1}
\and L. Izzo\inst{1,2,3} }

\institute{Dipartimento di Scienze Fisiche, Universit\`a di Napoli
"Federico II" and INFN Sez. di Napoli, Compl. Univ. Monte S.
Angelo, Ed. N, Via Cinthia, I-80126 Napoli, Italy \and ICRANet and
ICRA, Piazzale della Repubblica 10, I-65122 Pescara, Italy. \and
Dip. di Fisica, Universit\`a di Roma "La Sapienza", Piazzale Aldo
Moro 5, I-00185 Roma, Italy.}

\abstract {}{Relations connecting gamma ray burst quantities can
be used to constrain cosmographic parameters of the Hubble law  at
medium-high redshifts.}{We consider a sample of 27 gamma ray
bursts to construct the luminosity distance to  redshift relation
and derive the values of the parameters $q_0$, $j_0$, and $s_0$.
The analysis is compared with other methods in the literature. }{
gamma gay bursts, if calibrated by SNeIa, seem reliable as
distance indicators and give cosmographic parameters in agreement
with the $\Lambda$CDM model.}{}


\maketitle

\section{Introduction}
A class of very accurate standard candles, the supernovae Ia
(SNeIa), has been highly developed in the last two decades
\citep{Branch}; however, these objects are hardly detectable at
redshifts higher than $z$ = 1.7, so the study of more distant
regions of the Universe leads to the needing to implement more
powerful standard candles. The problem becomes particularly
crucial at intermediate redshift, $z$ = 6 - 7, where, up to now,
not very well - defined distance indicators are available.

In the last years, several efforts have been made in order to
implement gamma ray bursts (GRBs), the most powerful explosions in
the Universe, as  standard candles, and several interesting
results have  recently been achieved  (e.g. Amati et al 2008;
Basilakos $\&$ Perivolaropoulos 2008 and references therein).
Considering the standard model of such objects, the GRB phenomenon
should originate from the black hole formation and reach  huge
amounts of energy (up to $10^{54} erg$). These events are observed
at considerable distances, so there are several efforts to frame
them into the standard of cosmological distance ladder.

In the literature, several more - detailed models  give account
for the GRB formation, e.g. Meszaros 2006; Ruffini et al 2008,
but, up to now, none of them is intrinsically capable of
connecting all the observable quantities. For this reason, GRBs
cannot be used as standard candles. Despite  this shortcoming,
there are several observational correlations among the photometric
and  spectral properties of GRBs.  These features allow  use of
GRBs as distance indicators \citep{Schaefer}, even when they
cannot be fully ''enrolled" in the class of standard candles. In
particular, it is possible to connect the peak energy of GRBs,
$E_p$, with the isotropic energy released in the burst, $E_{iso}$,
and with the rest frame jet break - time of the afterglow optical
light curve, measured in days, $t_b$, \citep{Liang2}:

\begin{equation}\label{eq:no1}
 \log{E_{iso}}=a + b_1 \log{\frac{E_p (1+z)}{300keV}} + b_2 \log{\frac{t_b}{(1+z)1day}}
\end{equation}
where $a$ and  $b_i$, with $i=1,2$, are calibration constants.

Another interesting result is the relation given by Ghirlanda et
al. \citep{Ghirlanda}. It connects the peak energy $E_p$ with the
collimation-corrected energy, or the energy release of a GRB jet,
$E_{\gamma}$, where

\begin{equation}
 E_{\gamma} = ( 1 - \cos{\theta_{jet}} ) E_{iso},
\end{equation}
with $\theta_{jet}$ the jet opening angle, given by \citep{Sari}:

\begin{equation}
 \theta_{jet} = 0.163\left(\frac{t_b}{1 + z}\right)^{3/8}\left(\frac{n_0\eta_{\gamma}}{E_{iso,52}}\right)^{1/8},
\end{equation}
where $E_{iso,52} = E_{iso}/10^{52}$ ergs,  $n_0$ is the
circumburst particle density in 1 cm$^{-3}$, and $\eta_{\gamma}$
the radiative efficiency. The Ghirlanda et al. relation is

\begin{equation}\label{eq:no2}
 \log{E_{\gamma}} = a + b \log{\frac{E_p}{300 keV}},
\end{equation}
where $a$ and $b$ are two calibration constants.

These two relations are used the most   in constraining cosmology
due to their relatively small scatter,  interestingly very tight
in the Ghirlanda et al. one, and the sufficient number of data
points available.

In \citep{Schaefer},  an example of the  discrepancy between data
and theoretical curves is shown for these two relations. It is
worth noticing that the calibration of the above relations is
necessary to avoid the circularity problem: all the relations need
to be calibrated for every set of cosmological parameters. Indeed,
all GRB distances, obtained only in a photometric way, are
strictly dependent on the cosmological parameters since there is
no low-redshift  set of GRBs to achieve a cosmology-independent
calibration.

Recently, Liang et al. \citep{Liang}  present a calibration method
(Liang thereafter) for several GRB relations, included the above
relations (\ref{eq:no1}) and (\ref{eq:no2}), in a
cosmology-independent way using the SNeIa. In fact, the SNeIa are
very accurate standard candles, but their range is limited up to
$z \approx$ 1.7; hence, assuming that relations (\ref{eq:no1}) and
(\ref{eq:no2}) work at any $z$ and that, at the same redshift,
GRBs and SNeIa  have the same luminosity distance, it becomes
possible, in principle, to calibrate GRB relations at low
redshifts. The calibration parameters are shown in Table
\ref{table:no1}.
\begin{table}
\caption{Parameter values obtained by \citep{Liang}} 
\label{table:no1} 
\centering 
\begin{tabular}{c c c} 
\hline\hline 
Relation & a & b \\ 
\hline 
$E_{\gamma} - E_p$ & 52.26 $\pm$ 0.09 & 1.69 $\pm$ 0.11  \\ 
$E_{iso}-E_p - t_b$ & 52.83 $\pm$ 0.10 & 2.28 $\pm$ 0.30 \\
 &  & -1.07 $\pm$ 0.21  \\
\hline 
\end{tabular}
\end{table}
For the  $E_{iso}-E_p-t_b$ relation, the $b$-values in the first
line is $b_1$ and in the second line is $b_2$.

When our working-relations are calibrated  with the Liang method,
we can compute the luminosity distance $d_l$ from the well-known
relation between $d_l$ and the energy-flux ratio of the distance
indicators in consideration. Afterwards, we can use a formulation
given by Visser \citep{Visser1}, where the luminosity distance
$d_l$ is related to the cosmographic parameters \citep{Weinberg}
by means of a Taylor series expansion for the same $d_l$. Such an
analysis works very well at low and intermediate redshifts, since
very good classes of standard candles are available there.
Besides, it is useful to constrain alternative theories of
gravity, as shown in Capozziello et al. 2008. Since we are
calibrating GRBs by SNeIa (in the SNeIa redshift range, the $d_l$
Taylor series analysis works very well), the method could also be
extended  to the next step (intermediate-high redshifts) where
GRBs are expected to be suitable distance indicators. This working
hypothesis could be useful  in order to link low and high redshift
ranges and then fully probe $d_l$. However, it is clear that such
a Taylor expansion, derived for low redshifts, can be problematic
for fitting GRBs at high redshifts. Here, we consider it  a viable
methodological approach to link GRBs to SNeIa.

The aim of this work is  to achieve the cosmographic parameters
\citep{Weinberg} using the above GRB relations and then to test
the cosmological density parameters in a $\Lambda$CDM model. The
only assumption that we make here is that the Universe is
described by a Friedmann-Robertson-Walker geometry and the scale
factor of the universe $a(t)$ can be expanded in a Taylor series
(Sect.2). In Sect.3, after considering a sample of 27 GRBs, we use
 a best-fit analysis to derive  the cosmographic parameters discussed in
the previous section, adopting the so - called Chevallier,
Polarsky, Linder parameterization for the equation of state (EoS).
Discussion and conclusions are given in Sect.4.

\section{Cosmography}

The calibration  we want to achieve should be  cosmologically
model-independent; hence, applying the above relations to a  GRB
sample in a given $z$-range, we want to derive the related
cosmography. In particular, we want to obtain deceleration, jerk,
and snap parameters \citep{Visser1} and compare them with the
current values deduced by other methods and observations (see, for
example, Basilakos $\&$ Perivolaropoulos 2008; Capozziello et al
2008 and references therein).

Being  only related  to the derivatives of the scale factor allows
to fit the cosmographic parameters  versus the distance-redshift
relation without any \emph{a priori} assumption on the underlying
cosmological model but, this fails  at very high redshifts where
the Taylor expansion does not work yet.

To build a distance-redshift diagram,  one has to calculate the
luminosity distance for each GRB in a given sample. In our case
the luminosity distance is
\begin{equation}
\label{lum1}
 d_l = \left(\frac{E_ {iso}}{4\pi S_{bolo}'}\right)^{\frac{1}{2}},
\end{equation}
where $S_{bolo}' = S_{bolo}/(1+z)$ is  the bolometric fluence of
gamma rays in the burst, corrected with  respect to the rest
frame. The definition of $E_{iso}$ is different for each relation
used, therefore for the luminosity distance, we have

\begin{equation}
\label{lum2}
 d_l = \left[\frac{10^a \left(\frac{E_p(1+z)}{300 keV}\right)^{b_1}\left(\frac{t_b}{(1+z)1 day}\right)^{b_2}}{4 \pi S'_{bolo}}\right]^{1/2},
\end{equation}
adopting the Liang-Zhang relation, with $a$, $b_1$, and $b_2$
given in the Table \ref{table:no1}, and

\begin{equation}
\label{lum3}
d_l  = 7.575\frac{{(1 + z) a^{2/3} [E_p(1 + z)/100\,{\rm keV}]^{2b/3} }}{{(S_{bolo}
t_b )^{1/2} (n_0 \eta_\gamma  )^{1/6} }}{\rm{ Mpc}},
\end{equation}
for the Ghirlanda et al. relation, with $a$ and $b$ given in
\ref{table:no1}, \citep{xu}. Note that the former gives $d_l$ in
centimeters, therefore it  divides the result for the value of 1
parsec in cm, while the latter gives $d_l$ directly in Mpc.

The luminosity distance  can be connected to the Hubble series
\citep{Weinberg}. Expanding the Hubble law up to the fourth order
in redshift and considering the related luminosity distance, we
get \citep{Visser1}
\begin{eqnarray}
d_l(z) =  d_H z
\Bigg\{ 1 + {1\over2}\left[1-q_0\right] {z}
-{1\over6}\left[1-q_0-3q_0^2+j_0+ \frac{k \; d_H^2}{a_0^2} \right] z^2
\nonumber
\\
{+}
{1\over24}[
2-2q_0-15q_0^2-15q_0^3+5j_0(1+2q_0)+s_0
\nonumber
\\
+ \frac{2\; k \; d_H^2 \; (1+3q_0)}{a_0^2}]\; z^3 +
\mathcal{O}(z^4) \Bigg\}
\end{eqnarray}
where $d_H = c/H_0$ is the Hubble radius and where the
cosmographic parameters are defined as
\begin{equation}
H(t) = + {1\over a} \; {d a\over d t}\,,
\end{equation}
\begin{equation}
q(t) = - {1\over a} \; {d^2 a\over d t^2}  \;\left[ {1\over a} \;
{d a \over  d t}\right]^{-2}\,,
\end{equation}
\begin{equation}
j(t) = + {1\over a} \; {d^3 a \over d t^3}  \; \left[ {1\over a}
\; {d a \over  d t}\right]^{-3}\,,
\end{equation}
\begin{equation}
s(t) = + {1\over a} \; {d^4 a \over d t^4}  \; \left[ {1\over a}
\; {d a \over  d t}\right]^{-4}\,.
\end{equation}
They are usually referred to as the \emph{Hubble},
\emph{deceleration}, \emph{jerk}, and \emph{snap} parameters,
respectively. Their present values, which we denote with a
subscript $0$, may be used to characterize the evolutionary status
of the Universe. For instance, $q_0 < 0$ denotes an accelerated
expansion, while $j_0$ allows us to distinguish among different
accelerating models; a positive value of $j_0$ indicates that, in
the past, the acceleration reversed its sign. In this paper,
according to the WMAP observations,  we assume the value of the
Hubble constant $H_0 \simeq 70 \pm 2$ km/sec/Mpc \citep{WMAP}.

The cosmographic  parameters can be expressed in terms of the dark
energy density and the EoS. Following the prescriptions of the
\emph{Dark Energy Task Force}, \citep{DETF}, we use the
Chevallier-Polarski-Linder parameterization (CPL) for the EoS
\citep{CPL1,CPL2} and assume a spatially flat Universe filled with
dust matter and dark energy. The dimensionless Hubble parameter
$E(z) = H/H_0$ reads as
\begin{equation}
 E^2(z) = \Omega_M (1 + z)^3 + \Omega_{X} (1 + z)^{3(1 + w_0 + w_a)}e^{-\frac{3w_a z}{1+z}},
\end{equation}
with $\Omega_{X} = 1 - \Omega_M$ and $w_0$ and $w_a$ the CPL
parameterization for the EoS (see Chevallier et al 2001; Linder
2003; Capozziello et al 2008 for details). We can have $\Omega_{X}
\equiv \Omega_{\Lambda}$, with $\Lambda$ the cosmological
constant. Such a relation can be used to evaluate the cosmographic
parameters, obtaining

\begin{equation}\label{eq:no10}
q_0 = \frac{1}{2} + \frac{3}{2} (1 - \Omega_M) w_0 \ ,
\end{equation}

\begin{equation}\label{eq:no11}
j_0 = 1 + \frac{3}{2} (1 - \Omega_M) \left [ 3w_0 (1 + w_0) + w_a \right ]
\ ,
\end{equation}

\begin{eqnarray}\label{eq:no12}
s_0 & = & -\frac{7}{2} - \frac{33}{4} (1 - \Omega_M) w_a \nonumber \\ ~ & -
& \frac{9}{4} (1 - \Omega_M) \left [ 9 + (7 - \Omega_M) w_a \right ] w_0
\nonumber \\ ~ & - & \frac{9}{4} (1 - \Omega_M) (16 - 3\Omega_M) w_0^2 \nonumber \\
~ & - & \frac{27}{4} (1 - \Omega_M) (3 - \Omega_M) w_0^3 \,.
\end{eqnarray}

For a $\Lambda$CDM-universe, where $(w_0, w_a) = (-1, 0)$, it
becomes

\begin{equation}\label{eq:no3}
 q_0 = -1 + \frac{3}{2}\Omega_M;
\end{equation}
\begin{equation}\label{eq:no4}
 j_0 = 1;
\end{equation}
\begin{equation}\label{eq:no5}
 s_0 = 1 - \frac{9}{2}\Omega_M\,,
\end{equation}
that are the quantities which we are going to fit using a given
GRB sample.

\section{GRB data fitting}

Let us take  a GRB sample into account that satisfies the above
relations. Unfortunately only 27 GRBs have observed jet breaks in
the Schaefer sample \citep{Schaefer}.  The  observational
quantities of  GRBs to take into account, are listed in Table
\ref{table:no6}. The luminosity distance for each of the relations
is given by Eqs. (\ref{lum2}) and (\ref{lum3}), and then we obtain
a data distribution in the luminosity distance-redshift diagram
$d_l - z$. The  errors on the data are only of a photometric
nature and, in a first analysis,  we can exclude  errors on the
redshift. For each GRB, we assume $\eta_{\gamma} = 0.2$ and
$\sigma_{\eta} = 0$, \citep{Frail}.

\begin{table*}
\centering 
\caption{GRBs Data Sample} 
\label{table:no6} 
\begin{tabular}{c c c c c c c}
\hline\hline 
$GRB$ & $z$ & $E_p$ (keV) & $S_{bolo}$ (erg cm$^{-2}$) & $t_{jet}$ (days) & $\theta_{jet}$ (deg.) & $n_0$ $(cm^{-3})$ \\
(1) & (2) & (3) & (4) & (5) & (6) & (7)\\
\hline 
970508  & 0.84 & 389 $\pm$ 40 & 8.09E-6 $\pm$ 8.1E-7    &   25  $\pm$   5 & 23  $\pm$   3 &  3.0 $\pm$ 2.4 \\ 
970828  & 0.96 & 298 $\pm$ 30 & 1.23E-4 $\pm$   1.2E-5  &   2.2 $\pm$   0.4 &   5.91    $\pm$   0.79 & 3.0 $\pm$ 2.4  \\
980703  & 0.97 & 254 $\pm$ 25 & 2.83E-5 $\pm$   2.9E-6  &   3.4 $\pm$   0.5 &   11.02   $\pm$   0.8 & 28.0 $\pm$ 10   \\
990123  & 1.61 & 604 $\pm$ 60 & 3.11E-4 $\pm$   3.1E-5  &   2.04    $\pm$   0.46    &   3.98    $\pm$   0.57 &  3.0 $\pm$ 2.4 \\
990510  & 1.62 & 126 $\pm$ 10 & 2.85E-5 $\pm$   2.9E-6  &   1.6 $\pm$   0.2 &   3.74    $\pm$   0.28 &  0.29 $\pm$ 0.14 \\
990705  & 0.84 & 189 $\pm$ 15 & 1.34E-4 $\pm$   1.5E-5  &   1   $\pm$   0.2 &   4.78    $\pm$   0.66 &  3.0 $\pm$ 2.4 \\
990712  & 0.43 & 65 $\pm$ 10 & 1.19E-5  $\pm$   6.2E-7  &   1.6 $\pm$   0.2 &   9.47    $\pm$   1.2 & 3.0 $\pm$ 2.4  \\
991216  & 1.02 & 318 $\pm$ 30 & 2.48E-4 $\pm$   2.5E-5  &   1.2 $\pm$   0.4 &   4.44    $\pm$   0.7 & 4.7 $\pm$ 2.8  \\
010222  & 1.48 & 309 $\pm$ 12 & 2.45E-4 $\pm$   9.1E-6  &   0.93    $\pm$   0.1 &   3.03    $\pm$   0.14 &  3.0 $\pm$ 2.4 \\
011211  & 2.14 & 59 $\pm$ 8 & 9.20E-6   $\pm$   9.5E-7  &   1.56    $\pm$   0.16    &   5.38    $\pm$   0.66 &  3.0 $\pm$ 2.4 \\
020124  & 3.20 & 87 $\pm$ 18 & 1.14E-5  $\pm$   1.1E-6  &   3   $\pm$   0.4 &   5.07    $\pm$   0.64 &  3.0 $\pm$ 2.4 \\
020405  & 0.70 & 364 $\pm$ 90 & 1.10E-4 $\pm$   2.1E-6  &   1.67    $\pm$   0.52    &   6.27    $\pm$   1.03 &  3.0 $\pm$ 2.4 \\
020813  & 1.25 & 142 $\pm$ 14 & 1.59E-4 $\pm$   2.9E-6  &   0.43    $\pm$   0.06    &   2.8 $\pm$   0.36 &  3.0 $\pm$ 2.4 \\
021004  & 2.32 & 80 $\pm$ 53 & 3.61E-6  $\pm$   8.6E-7  &   4.74    $\pm$   0.5 &   8.47    $\pm$   1.06 & 30.0 $\pm$ 27.0  \\
030226  & 1.98 & 97 $\pm$ 27 & 8.33E-6  $\pm$   9.8E-7  &   1.04    $\pm$   0.12    &   4.71    $\pm$   0.58 &  3.0 $\pm$ 2.4 \\
030328  & 1.52 & 126 $\pm$ 14 & 6.14E-5 $\pm$   2.4E-6  &   0.8 $\pm$   0.1 &   3.58    $\pm$   0.45 &  3.0 $\pm$ 2.4 \\
030329  & 0.17 & 67.9 $\pm$ 2.3 & 2.31E-4   $\pm$   2.0E-6  &   0.5 $\pm$   0.1 &   5.69    $\pm$   0.5 &  1.0 $\pm$ 0.11 \\
030429  & 2.66 & 35 $\pm$ 12 & 1.13E-6  $\pm$   1.9E-7  &   1.77    $\pm$   1.0 &   6.3 $\pm$   1.52 &  3.0 $\pm$ 2.4 \\
041006  & 0.71 & 63 $\pm$ 12 & 1.75E-5  $\pm$   1.8E-6  &   0.16    $\pm$   0.04    &   2.79    $\pm$   0.41 & 3.0 $\pm$ 2.4  \\
050318  & 1.44 & 47 $\pm$ 15 & 3.46E-6  $\pm$   3.5E-7  &   0.21    $\pm$   0.07    &   3.65    $\pm$   0.5 & 3.0 $\pm$ 2.4  \\
050505  & 4.27 & 70 $\pm$ 23 & 6.20E-6  $\pm$   8.5E-7  &   0.21    $\pm$   0.04    &   3.0 $\pm$   0.8 & 3.0 $\pm$ 2.4  \\
050525  & 0.61 & 81.2 $\pm$ 1.4 & 2.59E-5   $\pm$   1.3E-6  &   0.28    $\pm$   0.12    &   4.04    $\pm$   0.8 & 3.0 $\pm$ 2.4  \\
050904  & 6.29 & 436 $\pm$ 200 & 2.0E-5 $\pm$   2E-6    &   2.6 $\pm$   1   &   8   $\pm$   1 &  3.0 $\pm$ 2.4 \\
051022  & 0.80 & 510 $\pm$ 22 & 3.40E-4 $\pm$   1.2E-5  &   2.9 $\pm$   0.2 &   4.4 $\pm$   0.1 & 3.0 $\pm$ 2.4  \\
060124  & 2.30 & 237 $\pm$ 76 & 3.37E-5 $\pm$   3.4E-6  &   1.2 $\pm$       &   3.72    $\pm$   0.15 & 3.0 $\pm$ 2.4  \\
060210  & 3.91 & 149 $\pm$ 35 & 1.94E-5 $\pm$   1.2E-6  &   0.33    $\pm$   0.08    &   1.9 $\pm$   0.17 &  3.0 $\pm$ 2.4 \\
060526  & 3.21 & 25 $\pm$ 5 & 1.17E-6   $\pm$   1.7E-7  &   1.27    $\pm$   0.35    &   4.7 $\pm$   1 & 3.0 $\pm$ 2.4  \\
\hline 
\end{tabular}
\\
\hspace{1mm}
References: \citep{Jimenez}; \citep{Metzger}; \citep{Djorgovski}; \citep{Kulkarni}; \citep{Israel}; \citep{Bjornsson}; \citep{Li}
\end{table*}

Another version of the Hubble series can be used  to improve the
data fit. If we consider the equation for the distance modulus,
\begin{equation}
 \mu = 25 + \frac{5}{ln(10)}ln[d_l/(1 Mpc)] + 25,
\end{equation}
and substitute the equation for $d_l$, we obtain a logarithmic
version of the Hubble series:
\begin{eqnarray}\label{log}
\ln[d_l / (z Mpc)] = \ln(d_H/Mpc) - \frac{1}{2}[-1+q_0] z
\nonumber
\\
+ \frac{1}{24} [-3 + 10q_0 + 9q_0^2 - 4(j_0 + 1 + \frac{kd_H^2}{a_0^2})]z^2
\nonumber
\\
+ \frac{1}{24} [4q_0(j_0 + 1 + {k d_H^2}{a_0^2}) + 5 - 9q_0 - 16q_0^2 - 10q_0^3
\nonumber
\\
+ j_0(7 + 4q_0) + s_0]z^3 + \mathcal{O}(z^4)\,.
\end{eqnarray}

This logarithmic version shows  the advantage when there is no
need to transform the uncertainties on the distance modulus. With
these considerations in mind, we perform a  polynomial
least-squares fit for each relation of the data assuming  Taylor
series polynomials, both in distance and in logarithmic distance.
We stop at order $n = 3$ both for the polynomial fit and for the
logarithmic fit. In the latter case, we obtain an estimate of the
snap parameter. Note that we are using least squares since, in
absence of any better data-fitting procedure, this is the standard
procedure  when assuming Gaussian distributed uncertainties.

The truncated polynomial used in the fits has the form
\begin{equation}
 d(z) = \sum_{i=1}^3 a_i z^i,
\end{equation}
 and
\begin{equation}
 \ln[d(z)/(z Mpc)] = \sum_{i=1}^3 b_i z^i
\end{equation}
for the logarithmic fit. In the latter case, the Hubble constant
enters as the $i=1$ component of the fit. As stated above, we use
$H_0$ as a constraint (a \emph{prior}).

The fits can be used to estimate the  deceleration and the jerk
parameters.  The logarithmic fit is better for estimating the snap
parameter through the values of the coefficients $a_i$ and $b_i$
and  their statistical uncertainties. The statistical
uncertainties on $q_0$ are linearly related to the statistical
uncertainties on the parameter $b_1$, while the statistical
uncertainties on $j_0$ and  $s_0$ depend  non -linearly on $q_0$
and its statistical uncertainty. It is worth noticing the
combination $j_0 + kd_H^2/a_0^2$, which is a well-known degeneracy
in Eq.(\ref{log}) \citep{Weinberg}. It means that we cannot
determine $j_0$ and $\Omega = 1 + kd_H^2/a_0^2$ separately, but we
need an independent determination of $\Omega$  to estimate the
value of the jerk parameter.

The results of the fits are presented  in Table \ref{table:no3}.
and all of them include the error on the data. For the calculation
of the uncertainties on $d_l$, we have followed the procedure
discussed in Xu et al. \citep{xu}. For example, the fractional
uncertainties on $d_l$ in the Ghirlanda et al. relation, without
the small angle approximation for $\theta_{jet}$ \citep{Sari}, are
given by

\begin{eqnarray}
 \left( {\frac{{\sigma _{d_L } }}{{d_L }}} \right)^2  & = & \frac{1}{4}\left[
 \left( {\frac{{\sigma _{S_{bolo}  } }}{{S_{bolo}  }}} \right)^2  \right] +\frac{1}{4}\frac{{1 }}{{(1 - \sqrt
   {C_{\theta } } )^2 }}\left[ \left( {\frac{{\sigma _a }}{a}}\right)^2 \right. \nonumber \\ & &
   \left. + \left( {b\frac{{\sigma _{E_p } }}{{E_p }}} \right)^2 +
      \left( {b\frac{{\sigma _b }}{b}\ln \frac{E_p}{100}} \right)^2  \right]+ \frac{1}{4}\frac{C_{\theta }}{{(1 - \sqrt
   {C_{\theta } } )^2 }}  \nonumber \\ & &\times \left[ {\left( {\frac{{3\sigma _{t_b } }}{{t_b }}} \right)^2
    + \left( {\frac{{\sigma _{n_0 } }}{{n_0 }}} \right)^2}\right]\,,
\end{eqnarray}
where $C_{\theta} = [\theta \sin{\theta} / 8 - 8 \cos{\theta}]^2$.
This shows that the uncertainties on $d_l$ in the Ghirlanda et al.
relation are very high, due to the dependence on several
parameters. For this reason, in  Fig \ref{fig:no3}, the prediction
bounds are plotted  at a $68\%$ confidence level, instead of
$95\%$, as in LZ-relation.

\begin{table}
\caption{Results of the fits. LZ is for Liang-Zhang relation, GGL for the Ghirlandaet al. one} 
\label{table:no3} 
\centering 
\begin{tabular}{c c c c} 
\hline\hline 
Fit & $q_0$ & $j_0 + \Omega$ & $s_0$ \\ 
\hline 
$d_l(z)$ LZ & $-0.94 \pm 0.30$ & $2.71 \pm 1.1$ &  \\ 
$d_l(z)$ GGL & $-0.39 \pm 0.11$ & $2.52 \pm 1.33$ &   \\
$\ln [d_l/z]$ LZ & $-0.68 \pm 0.30$ & $0.021 \pm 1.07$ & $3.39 \pm 17.13$  \\
$\ln [d_l/z]$ GGL & $-0.78 \pm 0.20$ & $0.62 \pm 0.86$ & $8.32 \pm 12.16$ \\

\hline 
\end{tabular}
\end{table}

As  said above, only statistical  uncertainties have been
considered and   other kinds of errors (systematics of
cosmological inference, modelling errors and  ''historical"
biases, Visser 2007b) have been neglected. If we do not assume
$H_0$ as a constraint, the analysis  gives $H_0 = 56$ $km/s/Mpc$,
which means that the data sample needs to be improved with further
GRBs  to give more reliable results.

\begin{figure}
\includegraphics[width=10.4cm, height=8cm]{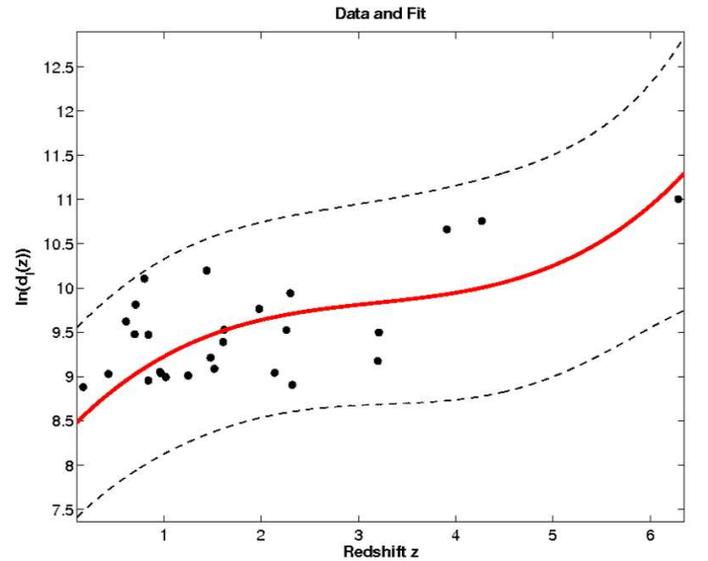}
\caption{Logarithmic version of the luminosity distance relation
versus redshift and the residuals of the $ln[d_l/zMpc]$ LZ fit. In this version of
the Hubble series, the discrepancy is the same at every $z$. The
dotted lines are  the bounds predicted at $95\%$  confidence
level. } \label{fig:no4}
\end{figure}

Another step would be to test  the goodness of the next fit
statistics using the MATLAB package. In particular, we  used the
R-square method: A value closer to 1 indicates a better fit. In
Table \ref{table:no5}, the results of  R-square are shown and  the
plots of the residuals of the fits are shown in Figs.
\ref{fig:no3}, \ref{fig:no4}.  For the logarithmic fit, the bad
value of the R-square is caused by the logarithm of the Hubble
series, which spreads a lot of the data on the $ln(d_l)$-axis. The
values $\ll 1$ for the logarithmic fits  are due to the
discrepancy of the data.

\begin{table}
\caption{Goodness of the fits with the R-square.} 
\label{table:no5} 
\centering 
\begin{tabular}{c c c c} 
\hline\hline 
Fit & R-square \\ 
\hline 
$d_l(z)$ LZ & $0.9909$ \\ 
$d_l(z)$ GGL & $0.9977$  \\
$\ln [d_l/z Mpc]$ LZ & $0.4005$   \\
$\ln [d_l/z Mpc]$ GGL & $0.2929$ \\

\hline 
\end{tabular}
\end{table}

In summary, the results are in  quite good  agreement with the
$\Lambda$CDM model, giving a Universe model that accelerates in
the present epoch and that has undergone a decelerated phase in
the past. The signature of this past phase is related to the sign
change of the parameter $q_0$ and the positive value of the jerk
parameter, unless a positive value of the spatial curvature
constant $k$ is considered. However this occurrence is excluded by
the last observational results, which confirm a spatially flat
Universe \citep{WMAP}.

\subsection{The CPL parameterization to test the $\Lambda$CDM model}

As said, the cosmographic parameters may also be expressed in
terms of the dark energy density and EoS parameters. Starting from
the Friedmann equation, we obtain the Hubble parameter:
\begin{equation}
 H^2 = \left(\frac{8\pi G}{3}\right)\rho,
\end{equation}
where $\rho$ is the energy density. The continuity equation for
each cosmological component  is given by the Bianchi identity
\citep{Weinberg}:
\begin{equation}
 \frac{\dot{\rho}}{\rho} = -3H \left(1 + \frac{p}{\rho}\right) = -3H [ 1 + w(z)],
\end{equation}
where $p$ is the pressure of the component considered and $w(z) =
p/\rho$  the redshift-dependent EoS for each component. The dark
energy component responsible for the observed acceleration of the
universe must have a negative EoS, \citep{Riess, Allen}. To find
$w(z)_{DE}$, we can adopt  the CPL parameterization, \citep{CPL1,
CPL2}, where
\begin{equation}
 w(z)_{DE} = w_0 + w_a z \left(\frac{1}{1 + z}\right)
\end{equation}
with $w_0$ and $w_a$ two parameters that enter directly into the
equations for the cosmographic parameters,
(\ref{eq:no10}-\ref{eq:no12}). To test the $\Lambda$CDM model, we
have  assumed $(w_0,w_a)=(-1,0)$.  Conversely, such a case can be
generalized by deducing the value of cosmographic parameters
$(\Omega_M, w_0, w_a)$ from polynomial fits where GRB data are
considered. Adopting our GRB sample,  we obtained the following
values
\begin{equation}
 w_0 = -0.53 \pm 0.64 \qquad w_a = 0.59 \pm 0.77,
\end{equation}
which directly enter into the CPL parameterization. The errors on
the CPL parameters are directly connected with the errors on the
cosmographic parameters, as   easily seen from the system
(\ref{eq:no10}-\ref{eq:no12}). The values of $w_0$ and $w_a$
agree, within the errors, with the $\Lambda$CDM model, without
assuming constraints {\it a priori} on the cosmological model.

\section{Discussion and conclusions}

Starting from  some relations connecting the observable quantities
of GRBs, we have used a sample of 27 GRBs  to derive the
luminosity distance - redshift diagram of the Hubble law. The
relations  conveniently calibrated by SNeIa  to make them
independent of any cosmological models.

We have taken the  Hubble law into account in the Taylor series
form, assuming the luminosity distance $d_l$ as a redshift
function whose coefficients are combinations of the cosmographic
parameters $H_0$, $q_0$, $j_0$, and $s_0$. The aim was to evaluate
such parameters starting from the GRB data. A direct analysis of
the fits leads to the conclusion that, in the error range, the
SNeIa results can also be extended  at higher redshifts
\citep{Visser3}. Besides, such results agree with the $\Lambda$CDM
model according to Eqs.
(\ref{eq:no3}),(\ref{eq:no4}),(\ref{eq:no5}). In particular, the
value of the parameter  $q_0$ that we found is in agreement with
the observed $\Omega_M$ (see Table \ref{table:no4}).

\begin{figure}
\includegraphics[width=10.4cm, height=8cm]{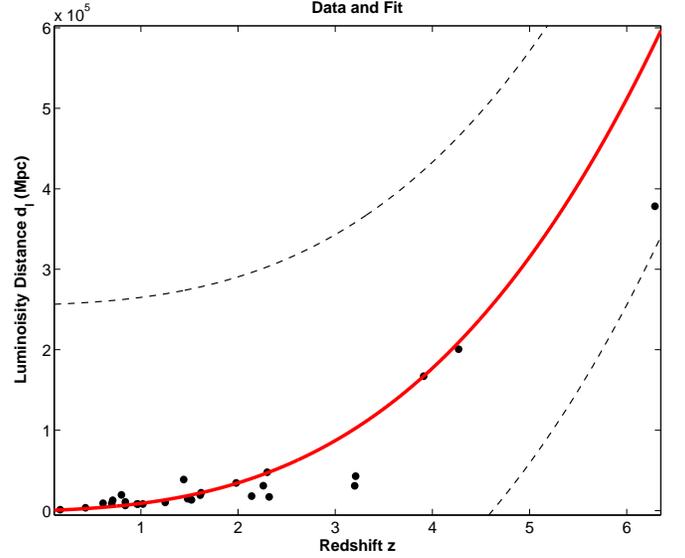}
\caption{Luminosity distance - redshift diagram and the residuals
of the $d_l(z)$ GGL fit. Note the discrepancy at high-$z$. The dotted lines
are  the bounds predicted at $68\%$ confidence level.}
\label{fig:no3}
\end{figure}

\begin{table}
\caption{Cosmological density parameters} 
\label{table:no4} 
\centering 
\begin{tabular}{c c c} 
\hline\hline 
Fit & $\Omega_M$ & $\Omega_{\Lambda}$ \\ 
\hline 
$d_l(z)$ LZ & $0.04 \pm 0.03$ & $0.65 \pm 0.73$  \\ 
$d_l(z)$ GGL & $0.46 \pm 0.43$ & $0.54 \pm 2.82$  \\
$\ln [d_l/z Mpc]$ LZ & $0.37 \pm 0.31$ & $0.63 \pm 1.13$ \\
$\ln [d_l/z Mpc]$ GGL & $0.28 \pm 0.30$ & $0.72 \pm 1.09$ \\
\hline 
\end{tabular}
\end{table}

However, the sample  we used is quite poor at high redshifts and,
in some sense, this justifies the use of the method of Taylor
series which works very well at low redshifts.  In particular, at
$z > 6$, we only have  one GRB, GRB050904 (see Fig.
\ref{fig:no3}). This GRB is very important in the fit results
because it affects the trend of the fits. For this reason we need
some richer sample at medium-high redshifts  to constrain  the
results better. However, if we had richer samples at high
redshifts, the Taylor series analysis would fail to constrain
cosmological models since an exact, and not approximated, $d_l(z)$
expression is needed in that case. The best constraint, however,
would be an absolute relation between several, GRB observables
which would make the GRBs a powerful standard candle at
intermediate-high redshift.

Considering these preliminary results, it seems that cosmography
by GRBs could be a useful tool in constraining self-consistent
cosmological models even if, up to now,  GRBs are not standard
candles in the proper sense.

\paragraph{Acknowledgements.}
We thank the referee for the useful suggestions that improved the
paper.


\begin{thebibliography}{99}

\bibitem[Albrecht et al 2006]{DETF}
Albrecht, A., et al, 2006, arXiv:astro-ph/0609591

\bibitem[Allen et al. 2004]{Allen}
Allen, S. W., et al, 2004, MNRAS, 353, 457

\bibitem[Amati et al 2008]{Amati}
Amati, L., et al, 2008, arXiv:astro-ph/0805.0377

\bibitem[Basilakos $\&$ Perivolaropoulos 2008]{Basilakos}
Basilakos, S., $\&$ Perivolaropoulos, L., 2008, arXiv: 0805.0875
[astro\,-\,ph]

\bibitem[Bjornsson et al. 2001]{Bjornsson}
Bjornsson, G., et al. 2001, ApJ, 552, L121

\bibitem[Branch $\&$ Tammann 1992]{Branch}
Branch, D., $\&$ Tammann, G. A. 1992, Ann.
Rev. Astron. Astrophys., 30, 359

\bibitem[Capozziello et al 2008]{cosmography}
Capozziello, S., et al, 2008, to appear in PRD, arXiv:
0802.1583[astro\,-\,ph]

\bibitem[Chevallier et al 2001]{CPL1}
Chevallier, M., Polarski, D., 2001, Int. J. Mod. Phys. D., 10, 213

\bibitem[Linder 2003]{CPL2}
Linder, E. V., 2003, Phys. Rev. Lett., \textbf{90} 091301

\bibitem[Djorgovski et al. 1999]{Djorgovski}
Djorgovski, S. G., Kulkarni, S. R., Bloom, J. S., Frail, D., Chaffee, F., $\&$
Goodrich, R. 1999b, GCN Circ. 189, http://gcn.gsfc.nasa.gov/gcn /gcn3/189
.gcn3

\bibitem[Frail et al. 2001]{Frail}
Frail, D. A., et al., 2001, ApJ, 562, L55

\bibitem[Ghirlanda et al. 2004]{Ghirlanda}
Ghirlanda G., Ghisellini G., $\&$ Lazzati D., 2004 ApJ, 616, 331

\bibitem[Israel et al. 1999]{Israel}
Israel, G., et al. 1999, A$\&$A, 348, L5

\bibitem[Jimenez et al. 2001]{Jimenez}
Jimenez, R., Band, D., $\&$ Piran, T., 2001, ApJ, 561, 171

\bibitem[Komatsu et al 2008]{WMAP}
Komatsu, E. et al, 2008, arXiv: astro-ph/0803.0547

\bibitem[Kulkarni et al. 1999]{Kulkarni}
Kulkarni, S. R. et al., 1999, Nature, 398, 389

\bibitem[Li et al.]{Li}
Li H. et al., 2008, Apj, 680, 92

\bibitem[Liang et al 2008]{Liang}
Liang, N., et al, 2008, arXiv: astro-ph/0802.4262

\bibitem[Liang $\&$ Zhang 2005]{Liang2}
Liang, E., $\&$ Zhang B., 2005, ApJ, 633, 611

\bibitem[Meszaros 2006]{Meszaros}
Meszaros, P., 2006, Rept. Prog. Phys., \textbf{69} 2259

\bibitem[Metzger et al. 1997]{Metzger}
Metzger, M. R., Djorgovski, S. G., Kulkarni, S. R., Steidel, C. C., Adelberger,
K. L., Frail, D. A., Costa, E., $\&$ Frontera, F. 1997, Nature, 387, 878

\bibitem[Qi et al.]{Qi}
Qi S., Wang F. $\&$ Lu T., 2008, arXiv: astro-ph/0803.4304

\bibitem[Riess et al. 2004]{Riess}
Riess, A. G., et al., 2004, ApJ, 607, 665

\bibitem[Ruffini et al 2008]{Ruffini}
Ruffini, R., et al, 2008, \emph{Gamma Ray Bursts. Proceedings XI Marcel Grossmann Meeting},
(World Scientific, Singapore)

\bibitem[Sari et al. 1999]{Sari}
Sari, R., Piran, T., $\&$ Halpern, J. P., 1999, ApJ, 519, L17

\bibitem[Schaefer 2007]{Schaefer}
 Schaefer, B. E., 2007, ApJ, 660, 16

\bibitem[Visser 2004]{Visser1}
Visser, M., 2004, Class. Quant. Grav., \textbf{21}, 2603

\bibitem[Visser 2007a]{Visser2}
Visser, M., $\&$ Catto$\ddot{e}$n, C., 2007a, Class. Quant. Grav., \textbf{24} 5985

\bibitem[Visser 2007b]{Visser3}
Visser, M., $\&$ Catto$\ddot{e}$n, C., 2007b, arXiv: gr-qc/0703122

\bibitem[Weinberg 1972]{Weinberg}
Weinberg, S., 1972, \emph{Gravitation and Cosmology: Principles
and applications of the general theory of relativity}, (Wiley, New
York).

\bibitem[Xu et al. 2005]{xu}
Xu D., Dai Z. G. $\&$ Liang E. W., 2005, Apj, 633, 603

\end{thebibliography}
\end{document}